\documentstyle[aps]{revtex}
\begin{document}
\title{Can all neurobiological processes be described by classical physics?}
\author{A.M. Lisewski}
\address{Max-Planck-Institut f\"ur Astrophysik,
Karl-Schwarzschild-Str. 1, 85740 Garching, Germany}
\date{\today}
\maketitle
\begin{abstract}
We discuss results recently given in an article by M. Tegmark \cite{mt99} where
he argues that neurons can be described appropriately by
pure classical physics. This letter is dedicated to the question
if this is really the case when the role of dissipation and noise -- the two
concurrent phenomena present in these biological structures --
is taken into account. We argue that dissipation and noise might
well be of quantum origin and give also a possible reason why 
neural dynamics is not classical.
\end{abstract}
In his recent work M. Tegmark (MT99) states that certain processes 
in biological neural networks, such as the firing of neurons, 
occur independently of coherent quantum effects. 
As a result these processes evolve according to classical physics  and 
are in addition subject to dissipation and noise, which account
for the interaction with some environment. The latter 
 involves "[...] all degrees of freedom that the observer
is not paying attention to." \cite{mt99}.
The author further concludes that 
 dissipation and noise should be added as extra classical entities.
In particular, dissipation turns out to be important, 
because its characteristic timescale becomes comparable to the typical
dynamical timescale of the system.

In this letter we discuss the role of dissipation and noise, arising
from the interactions between the system (here, biological neurons)
and  the environment in some more detail. We focus on the
question when dissipation and noise can considered to be classical
and when this is not the case at all. We think that a satisfactory answer
has not been given in MT99. Thus it remains an open question
whether neurobiological structures evolve independently
of quantum effects or not. 

\subsubsection*{Dissipation}

The interaction between system and environment produces
dissipation. This phenomenon reflects the possiblity that during time 
evolution accessible modes change into inaccessible modes belonging to
the environment. Our question here is what origin the dissipative
processes described in MT99 really have. To discuss this problem we utilize the
coarse-graining paradigm to distinguish between system and
environment. This means that an original (classical or quantum
mechanical) physical structure is subdivided into two parts which
eventually become the system and the environment
\footnote{The {\it original physical structure} is actually 
a system which does not have an environment. This means that
its temporal evolution does not depend on inaccessible
modes. In what follows, we call this particlar structure {\it subsystem}.}. Usually this
division is realized by introducing a coarse graining procedure, 
i.e. small spatial lenghtscales are filtered out appropriately (for
details, see \cite{rm96}). Then there are two basic possibilties:

{\it a}. Dissipation is essentially a classical phenomenon. Then dissipation
emerges after coarse graining a subsystem that is completely representable by 
classical physics. This kind of dissipation we call {\it classical dissipation}.

{\it b}. Dissipation is a quantum phenomenon. Then it
is a by-product of coarse graining a subsystem that
cannot be described by classical physics appropriately {\it and} the
resulting dissipation cannot 
be obtained via coarse graining a classical subsystem. 
This kind of dissipation we call {\it quantum dissipation}.\\
Examples for both kinds are known. For instance, consider a subsystem
that is represented by the Euler equation. Then the resulting (coarse
grained) system exhibits dissipation arising from unresolved scales \cite{mcc90}.\\
On the other hand in \cite{lis99} we showed that already
non-interacting quantum field theories reveal dissipative behavior
after being averaged over small spatial volumes.
However, it turns out that quantum dissipation is inherently different from
its classical counterpart. This is because the former acts
non-locally in time, i.e. the state at an arbitrary time $ t$
depends on the system's history between $ t - \tau_{\rm m}$ and $ t$, 
where $ \tau_{\rm m}$ denotes a non-vanishing memory time.
Contrary to that a classical subsystem that is completely determined
by local (in space and time) equations of motion would -- after spatial
coarse graining -- lead to a local evolution in time.
Thus the resulting equations would not involve a memory term and in
this situation $ \tau_{\rm m}$ would become inifinitesimal small.

\subsubsection*{Noise}

As a next point we briefly
discuss the nature of the noise term itself. Noise is generated
by (hidden) modes of the environment which -- during their
temporal evolution -- occasionally 
become part of the system. A detailed description of the noise
arising from the coarse graining procedure for quantum systems is 
given in \cite{rm96} as well as in \cite{lis99}. Again it is obvious to
introduce classical and quantum noise using the same kind of classification
as in the previous section. But then it is not immediately clear
whether there is a subtstantial difference between the latter and the
former. However, since noise always appears together with dissipation
it is natural to consider both -- noise and dissipation -- at the same
time and therefore to investigate their cumulative effects. For instance, noise in a system with quantum dissipation 
would be represented by a non-Markovian random process (because
of the dependence on the system's history) while on the other hand
classical dissipation would not change a given
Markovian character of noise.

\subsubsection*{Possible origins of dissipation and noise in neurons}

Now we ask if quantum noise and dissipation are relevant for
the processes taking place in neurons {\it despite} the fact that
(according to the results in MT99) coherent quantum effects over 
spatial distances become neglegible. The decoherence analysis
in MT99 can be discussed in the context of  arguments given
in \cite{lis99}. There we used the coarse graining paradigm to
show that near the classical limit additional terms appear
within the classical equations of motion. These terms account for 
quantum dissipation, quantum noise and for the quantum potential (or
Bohm-potential). In fact, the latter term is responsible
for non-local quantum effects in space. Thus the results in MT99 and 
in particular the expression
given for the off-diagonal elements of the system's density matrix
suggest that coherent quantum effects coming from the quantum potential
are unimportant. At the same time however, quantum dissipation and
quatum noise need not to be neglegible. To proof that they actually are, one
would additionally have to show that on timescales relevant
for the macrosystem these terms provide a very small contribution.
For example, one necessary condition would be the fact that the 
memory time, $ \tau_{\rm m}$, is much smaller than the dynamical time
of the system. At this point it is wothwhile to note that the 
dynamics of biological neurons involves some non-local temporal
behavior. The so-called 'neuron firing', which is the 
initial emission and following propagation of an action potential
(spike) along the axon,  prerequisites that 
incoming electric signals reach some certain threshold. More specificly,
excitatory postsynaptic signals propagate towards the axon-hillock
where they lead to a large propability for the emission of a spike
when the {\it sum} of these incoming signals 
within a short period of time exceeds a threshold \cite{g98}. This
process is called {\it temporal summation} and the aforementioned 'short period of time' 
is known to be of the order of tens of milliseconds \footnote{This
number is close to the dynamic timescale of a neuron being around
$10^{-3} - 10^{-1}$ s.}.\\
Thus signal processing in 
a neuron involves a memory effect. Facing this fact we ask the
obvious question if the memory time present in neural signal
processing has something to do with the memory time $ \tau_{\rm m}$
that results from a quantum mechanical description of the system. 
If $ \tau_{\rm m}$ turns out to be much smaller than memory timescales
observed in neurons then we would have to look for a classical theory that
explains the considered non-local effect. So far, it is not clear
how such a theory might look like and what the physical assumptions are that
form its basis.
But since coarse grained quantum mechanics provides very naturally a
non-local temporal
behavior it seems -- at least at the present stage -- reasonable not to preclude quantum 
physics from neurobiological processes.



%

%

%
%


\end{document}